# Uncertainty Analysis Of Future Projections Of Temperature, Precipitation, And Solar Radiation Under Global Warming Effect In Tehran, Iran

Ehsan Mosadegh [1,*], Iman Babaeian [2]

[1] Atmospheric Sciences Department, University of Nevada, Reno, NV 89557, USA.

[2] Climate Change Division, Climate Research Institute, Mashhad, Iran.

[*] Corresponding author
Email: emosadegh@nevada.unr.edu
Tel: +1 775-722-5325

## Abstract

In order to investigate the scope of uncertainty in projections of GCMs for Tehran province, a multi-model projection composed of 15 models is employed. The projected changes in minimum temperature, maximum temperature, precipitation, and solar radiation under the A1B scenario for Tehran province are investigated for 2011-2030, 2046-2065, and 2080-2099. GCM projections for the study region are downscaled by the LARS-WG5 model. Uncertainty among the projections is evaluated from three perspectives: large-scale climate scenarios downscaled values, and mean decadal changes. 15 GCMs unanimously project an increasing trend in the temperature for the study region. Also, uncertainty in the projections for the summer months is greater than projection uncertainty for other months. The mean absolute surface temperature increase for the three periods is projected to be about 0.8°C, 2.4°C, and 3.8°C in the summers, respectively. The uncertainty of the multi-model projections for precipitation in summer seasons, and the radiation in the springs and falls is higher than other seasons for the study region. Model projections indicate that for the three future periods and relative to their baseline period, springtime precipitation will decrease about 5%, 10%, and 20%, and springtime radiation will increase about 0.5%, 1.5%, and 3%, respectively. The projected mean decadal changes indicate an increase in temperature and radiation and a decrease in precipitation. Furthermore, the performance of the GCMs in simulating the baseline climate by the MOTP method does not indicate any distinct pattern among the GCMs for the study region.

## 1. Introduction

General Circulation Models (GCMs) are considered important tools for simulating the future global climate. These models are able to simulate different earth systems such as the atmosphere, oceans, and earth surface. However, projections of these models have low confidence and high uncertainty. Therefore, in climate change studies, Intergovernmental Panel on Climate Change (IPCC) recommends using multiple GCMs in climate simulations in order to consider the range of uncertainty in the projections (Parry 2007). Based on this suggestion, several studies have been conducted using the multi-model ensemble approach in climate change simulations (Caballero et al. 2007; Lopez et al. 2009; Beyene et al. 2010; Hay and McCabe 2010; Raje and Mujumdar 2010; Setegn et al. 2011). Adopting this approach, several studies have addressed the performance of the models in simulations. Giorgi and Mearns (2002) defined 'model performance' and 'model convergence' criteria as the two validation criteria for evaluating the skill of the GCMs in simulating climate variables in the present and future climates. By these two criteria, they evaluated the performance of nine GCM models in simulating mean seasonal temperature and precipitation for 22 regions on the Earth by employing the Reliability Ensemble Averaging (REA) method. This method reduces the range of uncertainty in the simulations by minimizing the effect of the outlier models (models with weak performance). These two criteria have been addressed in several studies and with different techniques (Perkins et al. 2007; Murphy et al. 2004; Tebaldi and Knutti 2007; Giorgi and Mearns 2003; Tebaldi et al. 2005; Knutti et al. 2010). Perkins et al. (2007) evaluated the performance of 14 GCM models based on their skill in simulating the baseline period in 12 regions in Australia. By using the daily simulations of precipitation, minimum temperature, and maximum temperature, and calculating the probability distribution functions of the observed and simulated variables, they ranked the GCM models based on their skill in simulating the climate variables over Australia. Wilby and Harris (2006) by proposing a probabilistic structure, evaluated the impact of different uncertainty sources in simulating annual flows. In their study, they used four GCMs, two downscaling techniques, two hydrological models, and two emission scenarios, and investigated the influence of each source on output results. Moreover, Dessai and Hulme (2007) by using two GCMs investigated present uncertainty in each part of the modeling process of simulating the climate change impact on water resources. Results of these two studies showed that if GCM variables are used as the inputs to the hydrological impact assessment models, the uncertainty of the GCM models will influence the final results, and the influence of GCMs will be greater than other sources of uncertainty. In another study, Semenov and Stratonovitch (2010) by using a combination of outputs of fifteen GCMs and the LARS-WG statistical downscaling model, investigated the uncertainty of these models in projecting the impact of climate change on the probability of heat stress during the flowering of wheat at four European locations. Rahmani and Zarghami (2013) investigate the performance of 15 GCMs in projecting the impact of climate change on temperature and precipitation in the northwestern region of Iran in the period of 2011-2030 by using the Ordered Weighted Averaging (OWA) approach. Likewise, Gohari et al. (2013) investigated the uncertainty in projecting the impact of climate change on gains and water demand by projecting the temperature and precipitation for the 2015-2044 period in the Zayanderood watershed by using multi-model ensemble scenarios. In both studies, the LARS-WG model was used to downscale the GCM outputs.

Climate change is projected to impact each component of the climate system in all regions of the world, but with different magnitude and confidence (IPCC, 2021; Mejia et al., 2019; Mosadegh et al., 2018; Mosadegh and Nolin, 2020). A few studies have investigated climate model projections of changes in climate variables over the 21st century for the Tehran region (Mosadegh and Babaeian, 2021), and have addressed the uncertainty of these projections (Mosadegh et al., 2013), and have investigated the potential impact of the projected changes in climate variables on other aspects of the environment such as air quality (Mosadegh et al., 2013; Mosadegh et al., 2021). In the present study, the range of changes of temperature, solar radiation, and precipitation under climate change in the present century has been investigated for an urban environment. For this purpose, with regard to the maximum use of GCMs, projections of 15

GCMs, which were downscaled by using the LARS-WG5 stochastic weather generator for three periods of 2011-2030, 2046-2065, and 2070-2099, have been included in this study in order to involve the range of uncertainty of the models in projections. GCM projections were investigated based on the performance of the models in simulating the present climate (historical period) and convergence of the simulations for the three future periods. Finally, GCM simulations were evaluated to identify which model would be more skillful over the study region.

## 2. Methodology

### 2.1. Study region

*Dushan Tappeh* synoptic station is located east of Tehran at 35° 42' N and 51° 20' E with a height of 1209.2 m above mean sea level. This station was selected due to having the longest observations in the study region, and the 1972-2009 period with 38-year daily observations was selected as the baseline period to synchronize with the historical period of GCMs. In preparing the LARS-WG input scenario file, daily observations of the baseline period are used. Therefore, observations of daily precipitation, daily minimum temperature, and daily maximum temperature for every single day of the baseline period were extracted from this station. Solar radiation was not available for the baseline period in the study region. Therefore, daily total sunshine hours were used as an alternative to the radiation input. LARS-WG uses a regression approach to convert daily sunshine hours to daily total solar radiation received by the Earth's surface at a site (Rietveld 1978).

### 2.2. GCMs and emission scenarios

Currently, the main and the most powerful tools for global climate change projections are the GCMs (Giorgi and Francisco 2001; Lane et al. 1999; Mitchell 2003). These models are based on physical concepts that are defined by mathematical equations which are solved on a three-dimensional grid on the Earth. 15 GCMs used in this study are a subset of the models which are used in the IPCC 4[th] assessment report which was published in 2007. All these models are the coupled Atmospheric-Oceanic models and most of them have been run for the 1960-2100 period. Features of these 15 models are presented in table 1. For simulating the future climate, the models consider an estimate of future emissions of greenhouse gases as their input. These estimates are called emission scenarios which consider a wide range of effective factors such as future human population, and economical and technological factors affecting emissions of greenhouse gases and aerosols. In this study, A1B emission scenarios from among SRES scenarios are used. This scenario is considered a moderate scenario and most of the GCMs have used this scenario in their climate simulations.

### 2.3. Downscaling

In this study, the LARS-WG5 model was used to downscale the GCM simulations. LARS-WG is a statistical downscaling model which is placed among stochastic weather generator tools. Weather generators can simulate statistical characteristics of local climate variables and can generate local-scale daily climate scenarios for a specific station (Wilby et al. 2004). First, in the calibration step, the model extracts the statistical characteristics of each variable from long-term baseline observations and, by using these characteristics, regenerates the probability distribution of the variables for the baseline period. Then, in the quality test step, the model compares the statistical characteristics of the generated and observed variables in each month assuming that the observations are a random sample from the existing sample which is considered as the true climate at the site. The comparison of the generated and observed data is done by performing some statistical tests such as t, F, and K-S tests, which are for comparing means, standard deviations, and probability distributions of the two data sets, respectively. Each test delivers test statistics and p-values for each month. The p-value is computed at the 0.05 significance level and the values above this level indicate the high probability of similarity of the two data sets. Eventually, in the final step, by using the calculated statistical characteristics of the local climate and GCM climate scenarios for the study grid, daily time series of each variable is generated for the desired period in the future (Semenov and Barrow 2002).

Table 1: Features of the GCMs from IPCC AR4 used in this study

| Country | Developer | GCM | Model acronym | Grid resolution | Emissios scenarios |
|---------|-----------|-----|---------------|-----------------|--------------------|
| Australia | Commonwealth Scientific and Industrial Research Organization | CSIRO-MK3.0 | CSMK3 | $1.9° \times 1.9°$ | SRA1B, SRB1 |
| Canada | Canadian Centre for Climate Modeling and Analysis | CGCM33.1 (T47) | CGMR | $2.8° \times 2.8°$ | SRA1B |
| China | Institute of Atmospheric Physics | FGOALS-g1.0 | FGOALS | $2.8° \times 2.8°$ | SRA1B, SRB1 |
| France | Centre National de Recherches Meteorologiques | CNRM-CM3 | CNCM3 | $1.9° \times 1.9°$ | SRA1B, SRA2 |
| France | Institute Pierre Simon Laplace | IPSL-CM4 | IPCM4 | $2.5° \times 3.75°$ | SRA1B, SRB1, SRA2 |
| Germany | Max-Planck Institute for Meteorology | ECHAM5-OM | MPEH5 | $1.9° \times 1.9°$ | SRA1B, SRB1, SRA2 |
| Japan | National Institute for Environmental Studies | MRI-CGCM2.3.2 | MIHR | $2.8° \times 2.8°$ | SRA1B, SRB1 |
| Norway | Bjerknes Centre for Climate Research | BCM2.0 | BCM2 | $1.9° \times 1.9°$ | SRA1B, SRB1 |
| Russia | Institute for Numerical Mathematics | INM-CM3.0 | INCM3 | $4° \times 5°$ | SRA1B, SRB1, SRA2 |
| UK | UK Meteorological Office | HadCM3 | HADCM3 | $2.5° \times 3.75°$ | SRA1B, SRB1, SRA2 |
| | | HadGEM1 | HADGEM | $1.3° \times 1.9°$ | SRA1B, SRA2 |
| USA | Geophysical Fluid Dynamics Lab | GFDL-CM2.1 | GFCM21 | $2.0° \times 2.5°$ | SRA1B, SRB1, SRA2 |
| USA | Goddard Institute for Space Studies | GISS-AOM | GIAOM | $3° \times 4°$ | SRA1B, SRB1 |
| USA | National Centre for Atmospheric Research | PCM | NCPCM | $2.8° \times 2.8°$ | SRA1B, SRB1 |
| | | CCSM3 | NCCCS | $1.4° \times 1.4°$ | SRA1B, SRB1, SRA2 |

### 2.4. Uncertainty in climate modeling

Studies show that GCMs provide a realistic projection of the future climate (Solomon 2007). The ability of GCMs to reproduce a broad range of climate attributes increases the confidence of climate scientists that key physical processes are included in climate change simulations (Doblas-Reyes et al. 2006; Palmer et al. 2004; Palmer et al. 2005). However, in climate change studies and in different parts of simulating the climate variables, different sources of uncertainty exist that can influence the final output of the study. Giorgi and Francisco (2001) showed that the main sources of uncertainty in climate change simulations on a regional scale stem from uncertainty in greenhouse gas emission scenarios, uncertainty in GCM simulations and their internal differences, and uncertainties in different methods of downscaling GCM simulations. Furthermore, Wilby and Harris (2006) and Dessai and Hulme (2007) showed that GCM models have the highest uncertainty compared to other components of hydrological impact assessment systems.

Intercomparison studies of GCMs indicate that climate variables are simulated with different degrees of accuracy by different models, and no single model delivers the best simulation for all variables and/or all regions (Lambert and Boer 2001; Gleckler et al. 2008). Furthermore, in climate change impact assessment studies, due to the influence of different sources of uncertainty on the output of the predicting system, projections do not have sufficient confidence. Therefore, it is recommended that for quantifying the range of uncertainty in the projections, the maximum number of available GCM models be used in simulations (Jones 2000; Wilby and Harris 2006). Lambert and Boer (2001) and Gleckler et al. (2008) showed that the multi-model mean of an ensemble simulation yields a closer estimate to reality. Giorgi and Mearns (2002) defined two "*reliability criteria*" for evaluating the performance of GCMs in simulations: 'model performance' criterion, which means that how well the models can simulate the baseline (present) climate, and 'model convergence' criterion, which investigates the convergence between the simulations of future climate across the models. By considering both these criteria, they investigated the accuracy of the models in simulating the mean seasonal changes of temperature and precipitation simulated by 9 GCMs over 22 regions of the world. They assigned weights to the calculated means of each GCM and calculated the weighted mean of the variables among the GCMs projections. However, Christensen (2010) showed that averaging the weighted simulations in a multi-model projection is not superior to an unweighted mean, and weighting the models incorporates a level of uncertainty to ensemble-based climate simulations.

### 2.4.1. Performance of GCMs in simulating mean climate

In order to investigate which GCM can generate closer simulations for the study region (performance criterion in GCMs), the Mean Observation Temperature Precipitation (MOTP) method was used to assign weights to each GCM based on the deviation of its baseline simulated mean temperature (or precipitation) from its mean observation values using the Eq. 1 (Gohari et al. 2013)

$$W_{ij} = \frac{\frac{1}{\Delta_{ij}}}{\sum_{j=1}^{n} \frac{1}{\Delta_{ij}}} \tag{1}$$

where $W_{ij}$ is the weight of the GCM $j$ in month $i$, $n$ is the number of total GCMs, and $\Delta_{ij}$ is the difference between the simulated mean temperature or precipitation and its corresponding observed value by GCM $j$ in a month $i$ in the baseline period. The GCM simulations over the study region for the observation period were obtained from the Canadian climate change database (http://www.cccsn.ec.gc.ca), and then the monthly means were calculated for each GCM in the baseline period.

### 2.4.2. Range of uncertainty in projections

In the present study, to investigate the convergence criterion, simulations of precipitation, minimum temperature, maximum temperature, and solar radiation by 15 GCMs are illustrated in box plots. In this method, all the simulations conducted by the 15 GCMs for the study region in each month are illustrated in a single figure. Therefore, it enables the reader to easily compare the present range of uncertainty and its change over the different periods in the simulations. In the figures, the bottom and top borders of each box demonstrate the first and third quartile of the simulations, and the bottom and top lines out of the boxes demonstrate the minimum and the maximum of the projections, respectively. The middle line on the boxes also demonstrates the median of the simulations. Moreover, the unweighted means of the simulations of the 15 GCMs are illustrated on each box. It is noteworthy that the box plot does not illustrate a certain range of changes in simulations, but shows the range of uncertainty from the perspective of the employed GCMs in the simulations.

## 3. Results and discussion

### 3.1. Validation of LARS-WG

Comparison of monthly means and standard deviations of the simulated and observed variables at the *Dushan Tappeh* station for the baseline period (1972-2009) is illustrated in fig 1. As the figure shows, in March and December, LARS-WG has the highest error in simulating the monthly means of precipitation, which is about 6.5 mm wet bias. In simulating the monthly means of minimum and maximum temperature, the difference is negligible. The greatest difference in simulating the mean minimum temperature is in November which is about 0.4 °C lower than the observations, and the maximum temperature is in October and December which is about 0.4 °C. In simulating the radiation, the greatest difference is in November and about 0.4 $Mj/m^2$.day lower than the observations.

Moreover, the generated monthly means were compared with their corresponding observations in the baseline period by the t-test. For every month, the test results were evaluated at the 0.05 confidence level. The test statistics and p-values for each variable in each month are given in table 2. The test results show that the simulated monthly means of all variables in all months are similar to their corresponding observations, and except for the monthly simulated radiation in November, simulations of all the four climate variables in all months are acceptable at the 0.05 confidence level.

### 3.2. Climate scenarios

Box plot was used to illustrate and analyze the range of uncertainty in the large-scale climate change scenarios projected by 15 GCMs for the study region. For plotting each box, projections of the 15 GCMs for each climate variable in each period were used. Moreover, to compare the changes in each period, the unweighted means of the 15 projections in each month are illustrated with bars on each box. In addition, to compare the future changes with the baseline period, diagrams of the long-term monthly means of each variable in the baseline period are shown on the boxes. Figs 2 and 3 illustrate the projected absolute changes in the minimum and maximum temperature relative to the baseline period in each month, respectively.

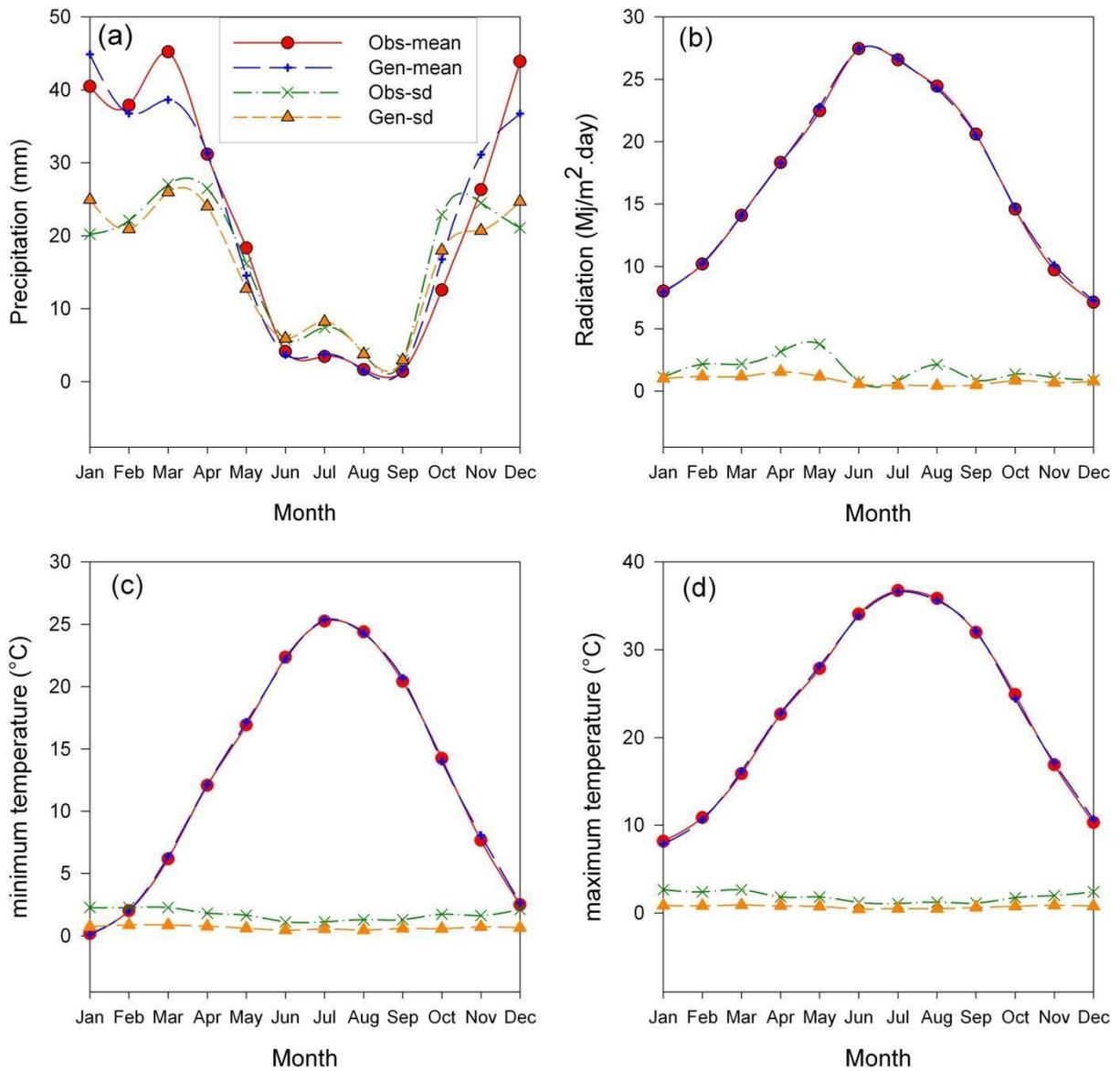

Fig 1: Comparison of monthly means and standard deviations of the simulated and observed variables at the *Dushan Tappeh* station in the baseline period (1972-2009) for (a) precipitation (mm), (b) radiation (Mj/m².day), (c) minimum temperature (°C), and (d) maximum temperature (°C).

As the boxes show, all GCMs unanimously project a temperature increase relative to the baseline period, but the increase is not uniform for all months. The rise in temperature in summers is projected to be higher than other months of the year, and therefore summers will get warmer than other months. Moreover, convergence in the projections for the summers is less than other months of the year, which means that the GCMs have more uncertainty in projecting the temperature in the summers than in other seasons. The convergence in the projections will also reduce in the long term in all months of the year. The mean absolute temperature increase under the A1B emission scenario is projected to be about 0.8 °C, 2.4 °C, and 3.8 °C in the summers, and about 0.5 °C, 1.5 °C, and 2.3 °C in the winters for the three periods, respectively.

Figs 4 and 5 illustrate the relative changes in the precipitation and solar radiation in percent, projected by 15

Table 2. Statistical tests and their p-values from LARS-WG

| Month | Precipitation | | | | Minimum temperature | | | | Maximum temperature | | | | Solar radiation | | | |
|---|---|---|---|---|---|---|---|---|---|---|---|---|---|---|---|---|
| | K-S | p-Value | t | p-Value | K-S | p-Value | t | p-Value | K-S | p-Value | t | p-Value | K-S | p-Value | t | p-Value |
| Jan | 0.058 | 1 | -0.87 | 0.385 | 0.106 | 0.9989 | 0.147 | 0.883 | 0.106 | 0.9989 | 0.697 | 0.488 | 0.044 | 1 | 0.363 | 0.718 |
| Feb | 0.04 | 1 | 0.234 | 0.816 | 0.053 | 1 | -0.08 | 0.936 | 0.105 | 0.9991 | 0.63 | 0.531 | 0.087 | 1 | -0.18 | 0.851 |
| Mar | 0.136 | 0.9743 | 1.124 | 0.264 | 0.053 | 1 | -0.56 | 0.573 | 0.053 | 1 | -0.76 | 0.45 | 0.087 | 1 | 0.016 | 0.987 |
| Apr | 0.069 | 1 | -0.03 | 0.971 | 0.106 | 0.9989 | -0.25 | 0.801 | 0.053 | 1 | -0.41 | 0.678 | 0.044 | 1 | 0.079 | 0.937 |
| May | 0.084 | 1 | 1.196 | 0.235 | 0.053 | 1 | -0.90 | 0.366 | 0.053 | 1 | -1.05 | 0.297 | 0.087 | 1 | -0.56 | 0.572 |
| Jun | 0.131 | 0.9824 | 0.333 | 0.74 | 0.106 | 0.9989 | 0.761 | 0.449 | 0.106 | 0.9989 | 0.977 | 0.331 | 0.044 | 1 | -0.02 | 0.978 |
| Jul | 0.117 | 0.9954 | -0.18 | 0.857 | 0.105 | 0.9991 | -0.67 | 0.503 | 0.105 | 0.9991 | 0.601 | 0.549 | 0.044 | 1 | -0.59 | 0.552 |
| Aug | 0.305 | 0.1932 | 0.255 | 0.799 | 0.106 | 0.9989 | 0.414 | 0.68 | 0.106 | 0.9989 | 1.032 | 0.305 | 0.131 | 0.9824 | 0.547 | 0.586 |
| Sep | 0.29 | 0.2415 | -0.47 | 0.638 | 0.053 | 1 | -1.18 | 0.238 | 0.053 | 1 | -0.81 | 0.419 | 0.087 | 1 | 0.778 | 0.439 |
| Oct | 0.057 | 1 | -0.95 | 0.342 | 0.105 | 0.9991 | 0.858 | 0.393 | 0.106 | 0.9989 | 1.509 | 0.135 | 0.044 | 1 | -0.55 | 0.584 |
| Nov | 0.088 | 1 | -0.97 | 0.331 | 0.053 | 1 | -1.51 | 0.134 | 0.053 | 1 | -0.89 | 0.375 | 0.044 | 1 | -1.99 | 0.05 |
| Dec | 0.042 | 1 | 1.412 | 0.162 | 0.105 | 0.9991 | -0.4 | 0.69 | 0.053 | 1 | -1.02 | 0.307 | 0.131 | 0.9824 | -0.71 | 0.477 |

GCMs for the three future periods relative to their baseline period for each month of the year, respectively. The figures show that the changes in the two variables have almost a reverse pattern so that the models project decreased precipitation and increased solar radiation for the springs, but increased precipitation and decreased radiation for the falls. The projected decrease in precipitation in the springs may be due to the reduced cloudiness and increased radiation in the season, which this trend will be reversed in the falls. Moreover, the projections state that the relative reduction in the precipitation reaches its peak in summers (June, July, and August), but because the precipitation and the cloudiness in this season are less than other seasons, no great changes in the amount of received radiation are projected for the summers.

Furthermore, the models project the least convergence and the most uncertainty in relative changes in the precipitation for the summers and as a reduction in precipitation. However, due to the small amount of precipitation in this season, this reduction will not have a great effect on the amount of precipitation received by the Earth's surface. This reduction will be more pronounced for the winters which have a high amount of precipitation. It is projected that, relative to the baseline period, the winter precipitation will be reduced on average by about 2%, 8%, and 15% from the first to the third period, respectively. In terms of solar radiation, the models show the highest uncertainty in the projected relative changes in radiation for the springs and then falls, and as an increase and a decrease in the radiation, respectively. It is projected that, relative to the baseline period, the spring radiation will be increased on average about 0.5%, 1.5%, and 3% from the first to the third period, respectively.

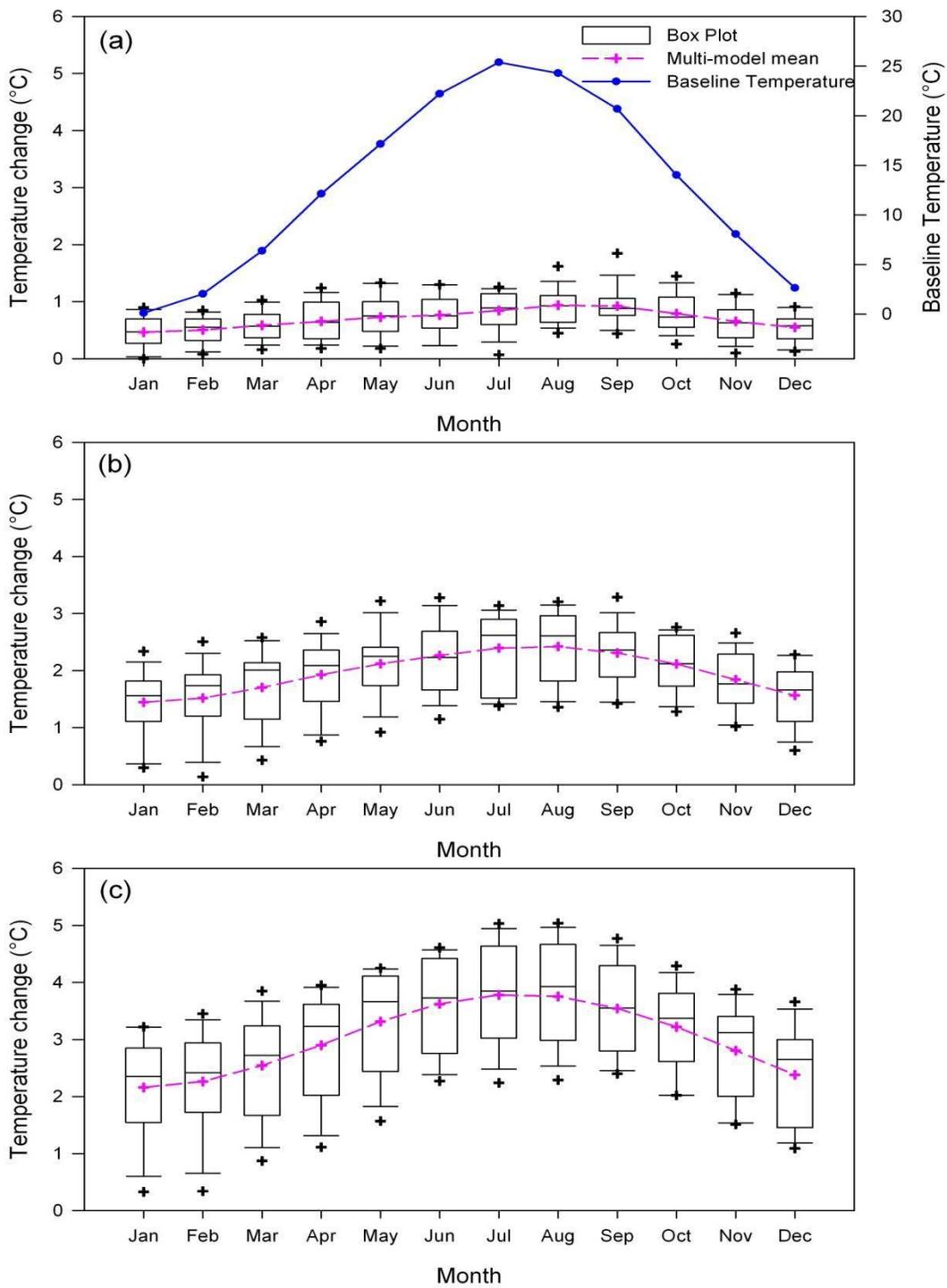

Fig 2: Projected absolute changes in the surface minimum temperature (°C) relative to the baseline period (1972-2009) at the *Dushan Tappeh* station for the three periods: (a) 2011-2030, (b) 2046-2065, and (c) 2080-2099.

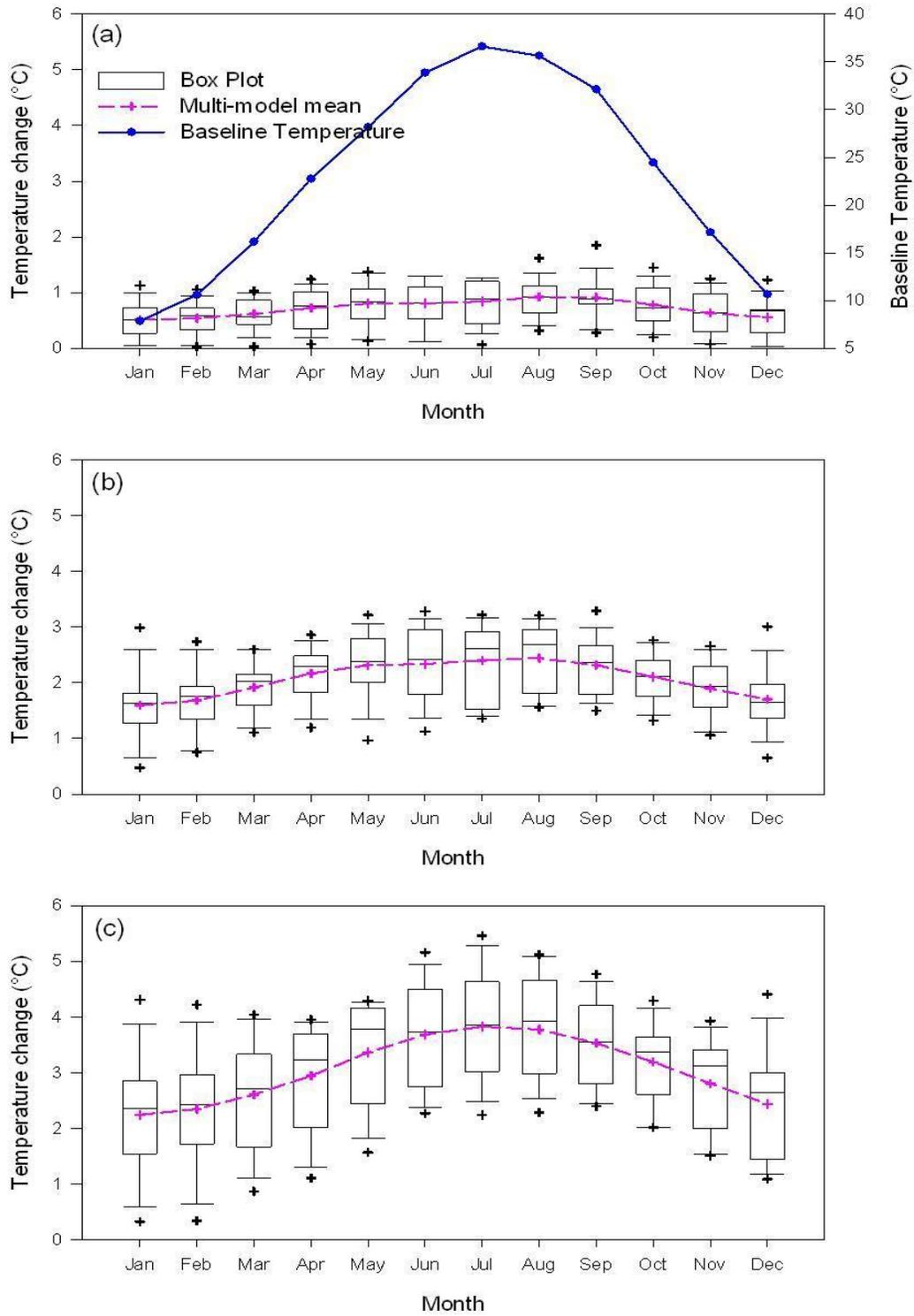

Fig 3: Projected absolute changes in the surface maximum temperature (°C) relative to the baseline period (1972-2009) at the *Dushan Tappeh* station for the three periods: (a) 2011-2030, (b) 2046-2065, and (c) 2080-2099.

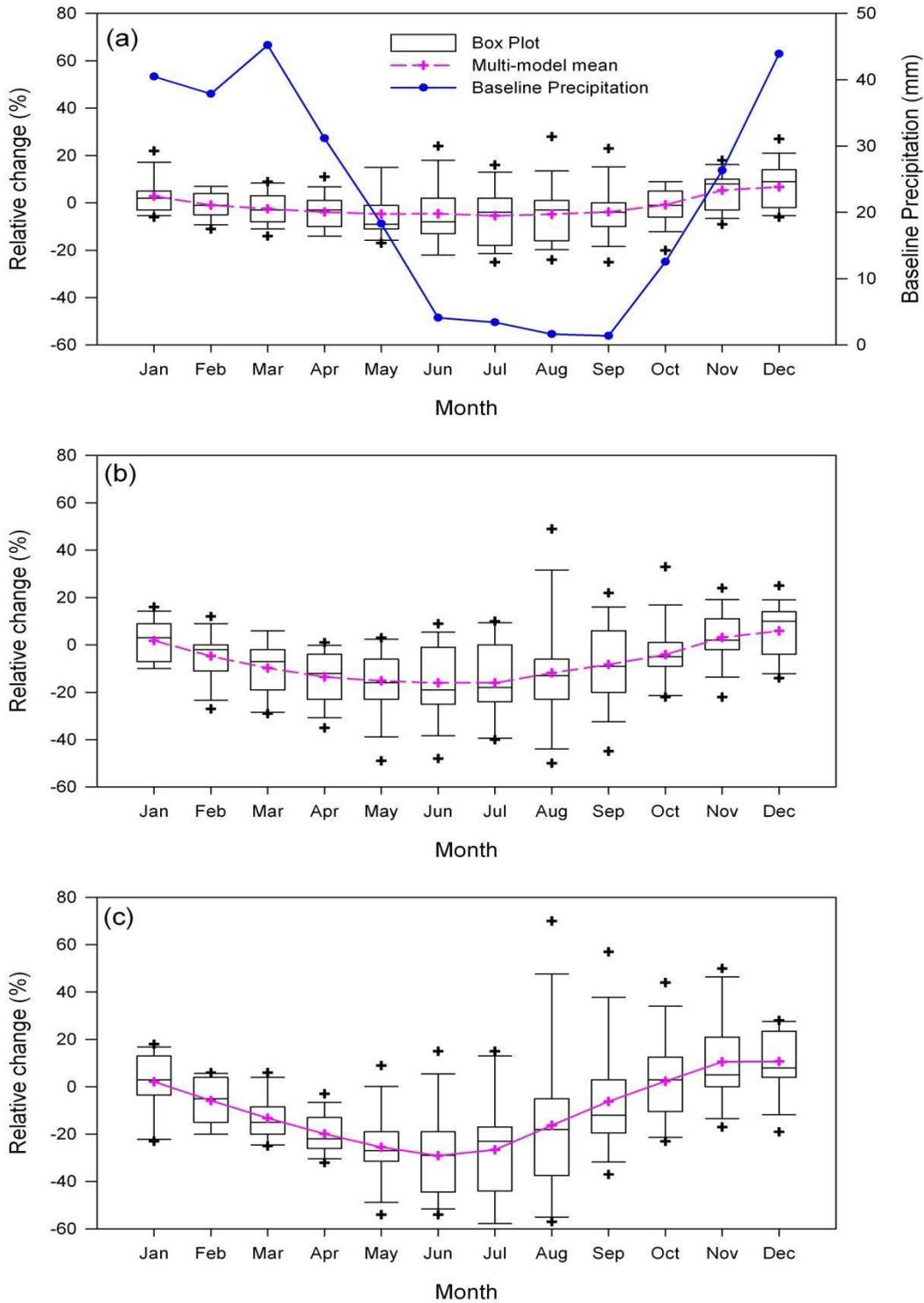

Fig 4. Projected relative changes in precipitation relative to the baseline period (1972-2009) at the *Dushan Tappeh* station for the three periods: (a) 2011-2030, (b) 2046-2065, and (c) 2080-2099.

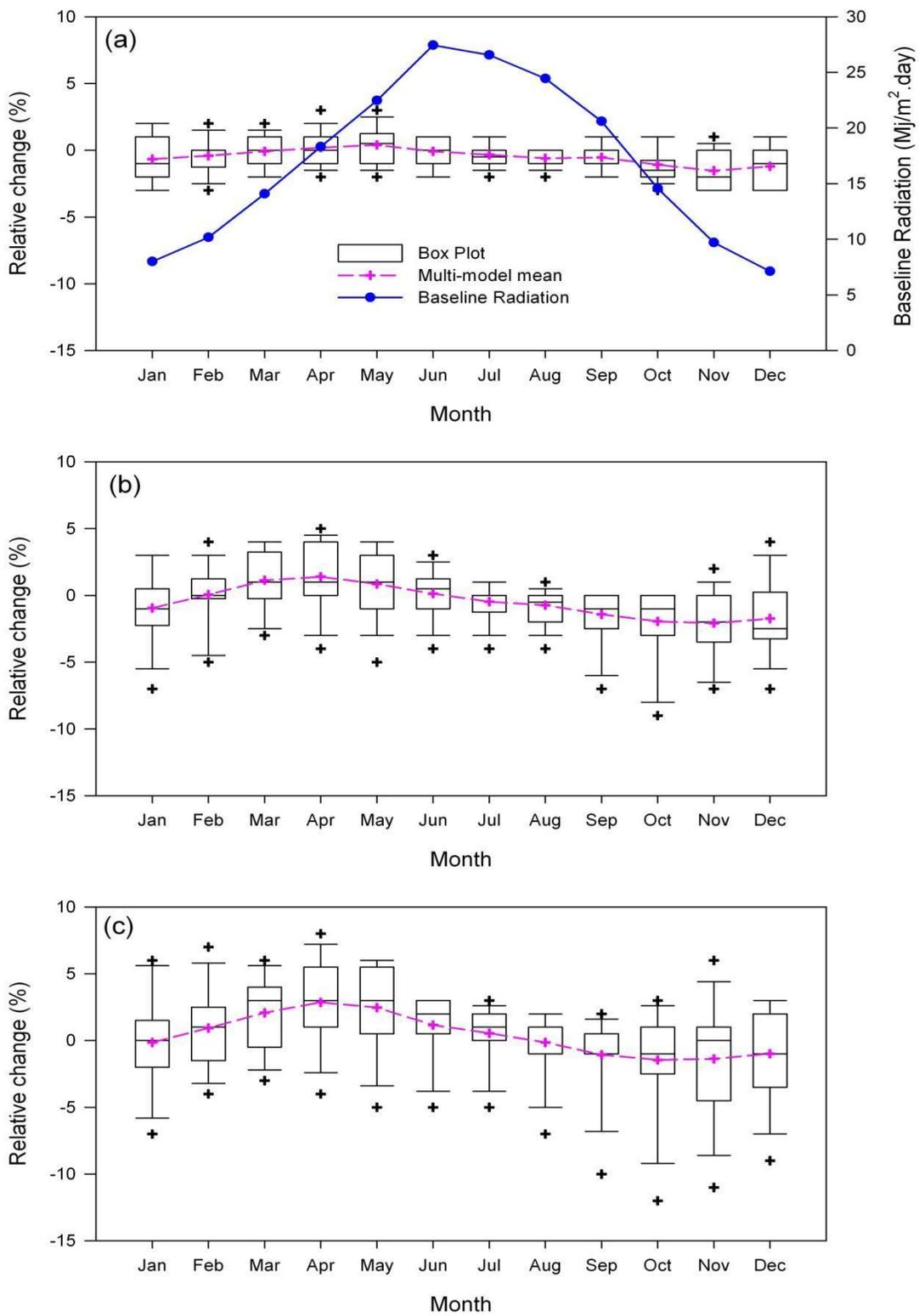

Fig 5. Projected relative changes in solar radiation relative to the baseline period (1972-2009) at the *Dushan Tappeh* station for the three periods: (a) 2011-2030, (b) 2046-2065, and (c) 2080-2099.

### 3.3. Downscaling GCM outputs

Projections of the 15 GCMs were downscaled for the study region and local-scale daily climate scenarios of the variables were obtained for each period. For investigating the convergence across the projections after the downscaling, all downscaled daily values of each variable in each month are gathered in one box. In order to compare the range of uncertainty entered to the projections after the downscaling, the changes of large-scale climate scenarios are also illustrated in this figure. Fig 6 shows this comparison for the maximum temperature. The small boxes on the left in each month are formed by adding the projected absolute temperature increase to the long-term observed monthly means of the baseline maximum temperature.

The boxes on the right in each month show the range of the downscaled daily maximum temperature, which are obtained from all downscaled daily values from LARS-WG for each month in each period. The baseline long-term monthly means of the maximum temperature are also illustrated under the boxes. The figures show that the changes in both diagrams follow the same trend. All diagrams show the temperature rise relative to the long-term baseline means, and this increase has grown by the end of the 21$^{st}$ century. However, a noticeable difference exists between the ranges of changes in the maximum temperature in the 2 sets of diagrams in each period. The downscaled values have a wider range compared to the large-scale projections. This increase in the range can be a result of the increase in climate fluctuations due to downscaling the variables from the grid-scale to the local scale (study station). Furthermore, a comparison of the figures in the three periods shows that the convergence in the projected large-scale climate scenarios (small boxes) in the summers is less than in other seasons, but after the downscaling, the convergence in the projections in the summers is more than other seasons.

### 3.4. Mean decadal changes of the climate variables for the *Dushan Tappeh* station

Changes in the climate variables in the future decades under the influence of climate change were also investigated from the mean decadal perspective. For each GCM, the decadal means of radiation, minimum and maximum temperature were obtained from averaging the downscaled daily values of each variable in each decade. For calculating the decadal mean of precipitation projected by each GCM, initially, monthly total precipitation was calculated from daily values and then the mean monthly total precipitation was calculated for each decade. Each box in fig 7 illustrates the projections of the 15 GCMs in comparison with the baseline long-term means of each variable. The ensemble projections indicate that the decadal changes of radiation, minimum, and maximum temperature will have an increasing trend, while the decadal changes of precipitation will have a decreasing trend in the coming decades. Decadal changes of the minimum and maximum temperature show a more distinct trend in comparison with the decadal changes of precipitation and radiation, and all GCMs unanimously project a noticeable growth in decadal changes of temperature. The figure indicates that the convergence in the projections reduces in the long term in all variables. Furthermore, the models project a reverse pattern for changes in radiation and precipitation in the future decades. The ensemble projections indicate that in the long term, precipitation will decrease, and in contrast, the radiation will increase in the study region.

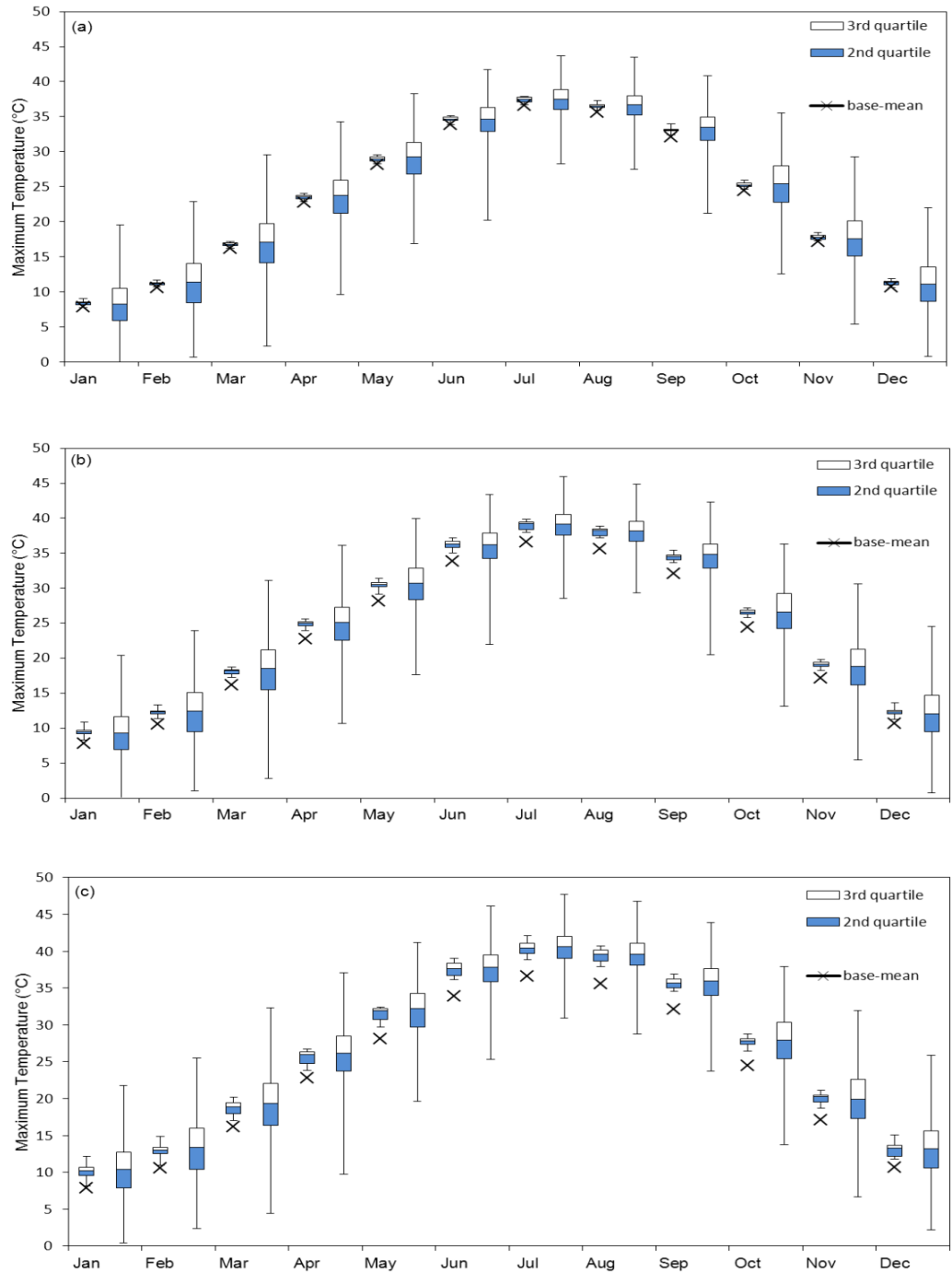

Fig 6. Comparison of convergence between daily values (downscaled by LARS-WG, large boxes) and large-scale climate scenarios (GCM outputs, small boxes) for the maximum temperature at the Dushan Tappeh station for the three periods: (a) 2011-2030, (b) 2046-2065, and (c) 2080-2099.

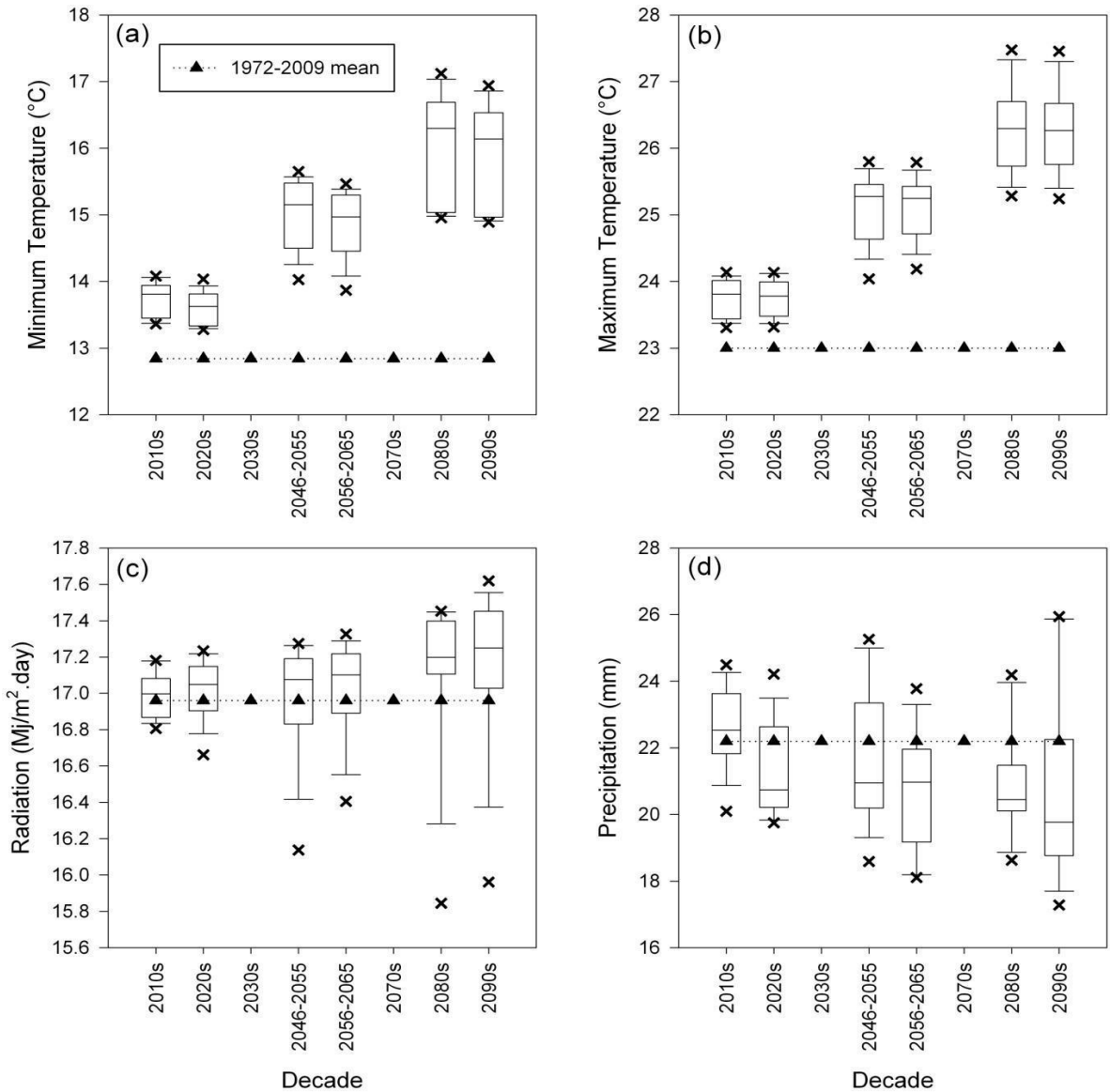

Fig 7. Projections of the mean decadal changes of (a) minimum temperature, (b) maximum temperature, (c) solar radiation, and (d) precipitation under the influence of climate change for the Dushan Tappeh station.

### 3.5. Weighting of the models

Studies indicate that the skills of GCMs in simulating all climate variables are not similar, and the accuracy of GCMs projections is different for different variables and/or regions (Gleckler et al. 2008). Therefore, the weighting of the projections by 15 GCMs for the study region was conducted by the MOTP method to investigate which GCM demonstrates better performance in simulating the baseline climate. Tables 3 shows that weighting the simulated monthly means of the variables do not show a distinct pattern, but it seems that in simulating the maximum

Table 3. Relative weights of the GCMs in simulating the monthly means of the variables: (a) daily maximum temperature (°C), (b) daily minimum temperature (°C), (c) daily mean temperature (°C), and (d) daily precipitation (mm/day).

(a)

| Month | BCM2 | CGMR | CNCM3 | CSMK3 | FGOALS | GFCM21 | GIAOM | HADCM3 | HADGEM | INCM3 | IPCM4 | MIHR | MPEH5 | NCCCSM | NCPCM |
|---|---|---|---|---|---|---|---|---|---|---|---|---|---|---|---|
| Jan | 0.02 | 0.05 | 0.03 | 0.23 | 0.03 | **0.33** | 0.09 | | | 0.11 | 0.07 | 0.04 | | | |
| Feb | 0.02 | 0.04 | 0.03 | **0.49** | 0.03 | 0.15 | 0.06 | | | 0.10 | 0.05 | 0.04 | | | |
| Mar | 0.02 | 0.05 | 0.04 | **0.57** | 0.04 | 0.08 | 0.04 | | | 0.08 | 0.04 | 0.05 | | | |
| Apr | 0.05 | 0.11 | 0.10 | **0.15** | 0.09 | 0.11 | 0.06 | | | 0.12 | 0.08 | 0.14 | | | |
| May | 0.03 | 0.12 | 0.06 | 0.08 | 0.07 | 0.07 | 0.03 | | | 0.06 | 0.04 | **0.43** | | | |
| Jun | 0.04 | 0.11 | 0.07 | 0.09 | 0.13 | 0.15 | 0.06 | | | 0.07 | 0.06 | **0.23** | | | |
| Jul | 0.04 | 0.04 | 0.11 | 0.06 | 0.15 | **0.34** | 0.06 | | | 0.05 | 0.04 | 0.12 | | | |
| Aug | 0.04 | 0.05 | 0.07 | 0.05 | 0.08 | 0.25 | 0.06 | | | 0.05 | 0.05 | **0.29** | | | |
| Sep | 0.04 | 0.15 | 0.07 | 0.05 | 0.06 | 0.15 | 0.07 | | | 0.06 | 0.06 | **0.29** | | | |
| Oct | 0.01 | **0.64** | 0.02 | 0.02 | 0.02 | 0.03 | 0.02 | | | 0.02 | 0.02 | 0.19 | | | |
| Nov | 0.05 | 0.11 | 0.07 | 0.11 | 0.06 | 0.11 | 0.09 | | | 0.09 | **0.18** | 0.12 | | | |
| Dec | 0.04 | 0.07 | 0.05 | 0.14 | 0.05 | 0.16 | 0.10 | | | 0.09 | **0.25** | 0.07 | | | |

(b)

| Month | BCM2 | CGMR | CNCM3 | CSMK3 | FGOALS | GFCM21 | GIAOM | HADCM3 | HADGEM | INCM3 | IPCM4 | MIHR | MPEH5 | NCCCSM | NCPCM |
|---|---|---|---|---|---|---|---|---|---|---|---|---|---|---|---|
| Jan | 0.02 | 0.02 | 0.03 | **0.53** | 0.02 | 0.05 | 0.09 | | | 0.03 | 0.19 | 0.03 | | | |
| Feb | 0.02 | 0.03 | 0.04 | **0.48** | 0.03 | 0.06 | 0.09 | | | 0.03 | 0.18 | 0.04 | | | |
| Mar | 0.05 | 0.06 | 0.07 | **0.28** | 0.07 | 0.09 | 0.11 | | | 0.06 | 0.14 | 0.07 | | | |
| Apr | 0.07 | 0.07 | 0.08 | **0.17** | 0.14 | 0.10 | 0.10 | | | 0.06 | 0.13 | 0.08 | | | |
| May | 0.07 | 0.07 | 0.09 | 0.14 | **0.17** | 0.09 | 0.09 | | | 0.06 | 0.13 | 0.09 | | | |
| Jun | 0.07 | 0.07 | 0.08 | 0.10 | **0.27** | 0.07 | 0.08 | | | 0.05 | 0.13 | 0.08 | | | |
| Jul | 0.06 | 0.07 | 0.07 | 0.07 | **0.38** | 0.06 | 0.07 | | | 0.04 | 0.12 | 0.07 | | | |
| Aug | 0.07 | 0.08 | 0.08 | 0.09 | **0.22** | 0.07 | 0.09 | | | 0.05 | 0.16 | 0.09 | | | |
| Sep | 0.07 | 0.08 | 0.09 | 0.09 | 0.15 | 0.08 | 0.10 | | | 0.05 | **0.21** | 0.09 | | | |
| Oct | 0.07 | 0.07 | 0.08 | 0.10 | 0.12 | 0.08 | 0.09 | | | 0.05 | **0.24** | 0.10 | | | |
| Nov | 0.05 | 0.05 | 0.05 | 0.09 | 0.06 | 0.06 | 0.07 | | | 0.04 | **0.48** | 0.06 | | | |
| Dec | 0.05 | 0.04 | 0.05 | 0.21 | 0.05 | 0.07 | 0.09 | | | 0.04 | **0.33** | 0.06 | | | |

(c)

| Month | BCM2 | CGMR | CNCM3 | CSMK3 | FGOALS | GFCM21 | GIAOM | HADCM3 | HADGEM | INCM3 | IPCM4 | MIHR | MPEH5 | NCCCSM | NCPCM |
|---|---|---|---|---|---|---|---|---|---|---|---|---|---|---|---|
| Jan | 0.02 | 0.03 | 0.02 | **0.40** | 0.02 | 0.08 | 0.07 | 0.03 | 0.03 | 0.04 | 0.12 | 0.03 | 0.04 | 0.04 | 0.02 |
| Feb | 0.02 | 0.03 | 0.03 | **0.44** | 0.03 | 0.08 | 0.07 | 0.03 | 0.03 | 0.05 | 0.07 | 0.04 | 0.04 | 0.03 | 0.02 |
| Mar | 0.03 | 0.05 | 0.05 | **0.31** | 0.05 | 0.08 | 0.06 | 0.04 | 0.04 | 0.06 | 0.07 | 0.06 | 0.04 | 0.04 | 0.03 |
| Apr | 0.05 | 0.07 | 0.07 | **0.13** | 0.09 | 0.08 | 0.06 | 0.04 | 0.04 | 0.06 | 0.09 | 0.08 | 0.05 | 0.04 | 0.04 |
| May | 0.05 | 0.08 | 0.08 | 0.11 | 0.12 | 0.08 | 0.06 | 0.04 | 0.03 | 0.06 | 0.08 | **0.13** | 0.04 | 0.04 | 0.03 |
| Jun | 0.05 | 0.15 | 0.06 | 0.08 | **0.16** | 0.07 | 0.06 | 0.03 | 0.02 | 0.04 | 0.07 | 0.12 | 0.03 | 0.03 | 0.02 |
| Jul | 0.04 | **0.30** | 0.05 | 0.05 | 0.19 | 0.06 | 0.05 | 0.02 | 0.02 | 0.03 | 0.05 | 0.11 | 0.02 | 0.02 | 0.02 |
| Aug | 0.05 | **0.21** | 0.07 | 0.06 | 0.13 | 0.07 | 0.07 | 0.02 | 0.02 | 0.04 | 0.08 | 0.12 | 0.03 | 0.03 | 0.02 |
| Sep | 0.06 | **0.14** | 0.07 | 0.06 | 0.10 | 0.08 | 0.08 | 0.03 | 0.03 | 0.05 | 0.11 | 0.10 | 0.03 | 0.03 | 0.03 |
| Oct | 0.05 | 0.10 | 0.06 | 0.08 | 0.07 | 0.08 | 0.07 | 0.04 | 0.04 | 0.05 | 0.12 | **0.13** | 0.04 | 0.04 | 0.03 |
| Nov | 0.05 | 0.06 | 0.05 | 0.09 | 0.05 | 0.07 | 0.07 | 0.04 | 0.05 | 0.05 | **0.22** | 0.08 | 0.05 | 0.05 | 0.03 |
| Dec | 0.02 | 0.02 | 0.02 | 0.08 | 0.02 | 0.05 | 0.04 | 0.02 | 0.03 | 0.03 | **0.56** | 0.03 | 0.03 | 0.03 | 0.02 |

(d)

| Month | BCM2 | CGMR | CNCM3 | CSMK3 | FGOALS | GFCM21 | GIAOM | HADCM3 | HADGEM | INCM3 | IPCM4 | MIHR | MPEH5 | NCCCSM | NCPCM |
|---|---|---|---|---|---|---|---|---|---|---|---|---|---|---|---|
| Jan | 0.07 | 0.02 | **0.01** | 0.01 | 0.02 | 0.01 | 0.02 | 0.01 | 0.08 | 0.01 | 0.19 | 0.04 | 0.01 | **0.48** | 0.01 |
| Feb | 0.01 | 0.04 | **0.74** | 0.01 | 0.03 | 0.03 | 0.01 | 0.01 | 0.01 | 0.01 | 0.03 | 0.03 | 0.01 | 0.02 | 0.01 |
| Mar | 0.01 | 0.15 | 0.01 | 0.02 | 0.13 | 0.09 | 0.01 | 0.03 | 0.02 | 0.02 | 0.06 | 0.07 | 0.03 | **0.33** | 0.01 |
| Apr | 0.00 | 0.01 | 0.00 | 0.02 | 0.01 | 0.01 | 0.00 | 0.05 | 0.00 | 0.22 | 0.06 | 0.08 | **0.26** | 0.22 | 0.04 |
| May | 0.00 | 0.01 | 0.01 | 0.03 | 0.02 | 0.14 | 0.01 | 0.06 | 0.01 | 0.02 | 0.19 | 0.11 | 0.06 | 0.17 | 0.17 |
| Jun | 0.00 | 0.02 | 0.00 | 0.10 | 0.01 | 0.07 | 0.02 | 0.08 | 0.02 | 0.01 | 0.05 | 0.09 | 0.01 | 0.14 | **0.39** |
| Jul | 0.01 | 0.18 | 0.01 | 0.04 | 0.14 | 0.03 | 0.07 | 0.14 | 0.01 | 0.01 | 0.05 | 0.04 | 0.01 | **0.19** | 0.05 |
| Aug | 0.01 | **0.44** | 0.00 | 0.04 | 0.09 | 0.05 | 0.05 | 0.13 | 0.00 | 0.01 | 0.05 | 0.05 | 0.01 | 0.02 | 0.07 |
| Sep | 0.01 | 0.04 | 0.00 | 0.01 | 0.03 | **0.47** | 0.01 | 0.02 | 0.00 | 0.01 | 0.04 | 0.09 | 0.01 | 0.01 | 0.23 |
| Oct | 0.02 | **0.17** | 0.01 | 0.04 | 0.10 | 0.13 | 0.01 | 0.09 | 0.01 | 0.07 | 0.05 | 0.15 | 0.03 | 0.04 | 0.08 |
| Nov | 0.00 | 0.02 | 0.02 | 0.07 | 0.09 | **0.22** | 0.02 | 0.05 | 0.02 | 0.16 | 0.16 | 0.07 | 0.04 | 0.03 | 0.03 |
| Dec | 0.08 | 0.04 | 0.03 | 0.03 | 0.04 | 0.03 | 0.07 | 0.02 | 0.08 | 0.03 | 0.06 | 0.07 | 0.03 | **0.38** | 0.02 |

temperature, MIHR, and CSMK3 show a better performance in the warm and cold months, respectively. In simulating the minimum temperature, CSMK3, FGOALS, and IPCM4 perform better in the simulations in the first 4 months, second 4 months, and the last 4 months of the year, respectively. In simulating the daily mean temperature, no distinct pattern can be seen among the simulations, and only the CSMK3 shows a better performance in simulating the daily mean temperature for the first 4 months of the year. In simulating the precipitation, similar to the daily mean temperature, no distinct pattern can be seen among the models, although it seems that NCCCSM shows a better performance in simulating the precipitation in some months for the study region.

**4. Conclusion**

In the present study, the uncertainty of 15 GCMs in projecting the changes in temperature, precipitation, and radiation for the present century under the influence of climate change was investigated for Tehran province. All models unanimously project an increase in temperature in all months, and this increase is projected with more uncertainty for summer seasons. The projections indicate a reverse pattern for the precipitation and radiation, showing a reduction in precipitation and an increase in the radiation are projected for the springs. Projections show that this pattern will reverse for the falls. Moreover, the convergence of the ensemble projections for precipitation in the summers, and the radiation in the springs and falls are less than other seasons for the study region. The uncertainty in the projections of the variables also grows in the long term (by the end of the 21$^{st}$ century), which indicates the weakness of the GCMs in long-term climate projections. Furthermore, decadal changes of the variables indicate that temperature and radiation will increase, and precipitation will decrease in the long term, and these changes are more pronounced for the temperature.

The projected changes in the climatic variables can influence the future urban environment of the study region. The projections indicate that Tehran will experience hotter summers in the future. This, together with the increased sunshine in the springs and summers, can influence temperature-related phenomena such as photochemical pollution and may degrade the future summertime air quality in the study region. Moreover, the projected reduction in winter and spring precipitation, which has the most precipitation among other seasons, may influence the local water resources and lead to water shortage in the study region. Therefore, the consequence of the projected changes in the future climate of the study region, which is considered as the first step among the steps of climate change impact assessment, can be investigated on other environmental aspects such as air quality and water resources in the study region to evaluate the scope of the impact of these changes on other resources and to devise a mitigation and adaptation strategy for the future decades. Therefore, future work should benefit greatly from techniques such as using dynamical-regional downscaling techniques, employing more local stations, and employing impact assessment models.

**Acknowledgment:**

The authors would like to give special thanks to Iran Meteorological Organization for providing meteorological data used in the current study. We are also grateful to the National Climatological Research Institute of Mashhad for its contribution to this paper.